\documentclass[11pt]{llncs}

\usepackage[english]{babel}
\usepackage{amsmath,amsfonts,latexsym}
\usepackage[all]{xy}

\newcommand{\N}{\mathbb N}
\newcommand{\Z}{\mathbb Z}

\begin{document}

\title{Advanced Software Protection Now}
\author{Diego Bendersky\inst{1,2}\and Ariel Futoransky\inst{1}\and Luciano 
Notarfrancesco\inst{1}\and Carlos Sarraute\inst{1,3}\and Ariel 
Waissbein\inst{1,3}}

\institute{Corelabs, Core Security Technologies; 
\and Departamento de Computaci\'on, FCEyN, Universidad de Buenos 
Aires (UBA), Argentina;
\and Departamento de Matem\'atica, FCEyN, UBA Argentina.}

\maketitle

\begin{abstract}
Software digital rights management is a pressing need for the
software development industry which remains, as no practical 
solutions have been acclamaimed succesful by the industry.
We introduce a novel software-protection method, 
fully implemented with today's technologies, that provides traitor
tracing and license enforcement and requires no 
additional hardware nor inter-connectivity.

Our work benefits from the use of secure triggers (\cite{FuKaSaWa:2003}),
a cryptographic primitive that is secure assuming the existence of an
{\sf ind-cpa} secure block cipher. Using our framework, developers 
may insert license checks and fingerprints,  and obfuscate the code 
using secure triggers. 
As a result, this rises the cost that software analysis tools have detect 
and modify protection mechanisms. Thus rising the complexity of cracking this
system.
\end{abstract}



\begin{section}{Introduction} 
Software piracy has troubled the computer industry,
producing millions of dollars of losses, and rising numerous scientific and 
technical problems of interest in computer security
(see, e.g., \cite{devanbu00software}, \cite{HachezEtal00}, \cite{WangEtal01}). 
Software is hardly sold, but it is typically licensed according
to policies defined by software license owners. 
Licensed software is executed within the licensed customers' computers and 
is expected to be run according to license policy.
For example, the license may establish that only users from an 
authorized IP address can use it, or that it can only run on a specific 
computer, or establishes an expiration date. 
However, the license owner does not have any technical 
warranties to enforce his policy, unless he uses a secure software 
protection system. The need for one such system 
remains after a long history of trials  (see, e.g., \cite{vanOorschot:2003}, 
\cite{devanbu00software}). 


\begin{subsection}{Background}
Software protection aims at enforcing license policy
through technical means, sometimes profiting from special-purpose 
hardware devices, and being supported by digital rights management 
legislations. 
Fingerprinting is roughly defined as the act of uniquely marking and
registering a build of the program allowing the license owner 
to trace back a copy of this build to its original licensee (\cite{chor94tracing}). 
Today, there is no agreement on how can license enforcement and
traitor tracing be implemented.
It is commonly acknowledged that given sufficient time (within human 
reach) and effort an attacker will crack any protection system.
As a result, software protection systems attempt to discourage crackers by making 
the cracking job a highly difficult (e.g., time-consuming) task. 

In the past, software protection solutions have tried to make 
(portions of) the software ``unavailable'' for inspection by
its users. 
Assuming available a procedure that provides this functionality,
we could construct a software protection system  by making license 
checks in this ``unavailable mode''  (because an attacker cannot 
thwart license checks he cannot access).
Among other things, software protection studies how to provide
this ``unavailability'' through code obfuscation 
(see, e.g., \cite{Hohl:1998}, \cite{sarmenta99}, \cite{sander98towards}).
Obfuscation, in software engineering, comprises the techniques used for 
preventing software analysis methods to produce qualitative results.
On the other side, reverse engineering (e.g., program comprehension; see,
e.g., \cite{ChikofskyCross:1990}), is the {\em de facto} software analysis 
discipline.

Regrettably, Barak {\em et alii} (\cite{barak01impossibility}) give a negative
result on obfuscation in the context of computational complexity, namely 
that obfuscation procedures can be defeated. 
More explicitly, they show that for every probabilistic polynomial-time Turing 
machine obfuscator, which receives as input a polynomial-time Turing 
machine (hereafter TM) and outputs a polynomial-time TM with the same 
functionality but such that only its input/output behavior is revealed by 
inspection, there exists a polynomial ``deobfuscator'' that learns more
than just the I/O behavior.

Notice that \cite{barak01impossibility} gives only an existential
result, and the construction of {\em deobfuscators} remains an
interesting problem.
Moreover, manual obfuscation is still possible, and the ``international 
obfuscated C code contest''  (\cite{iocc}) is a good example of this. 
It seems that the notion of automated TM obfuscation is too 
restrictive, and that a straight-forward computational complexity approach 
to software protection will not suffice to give a definitive answer (whether 
software protection is possible or not). 

Our thesis is that an hybrid approach, which combines computational 
complexity with software engineering, can be used to analyze a software
protection system with the necessary detail.
Evidence of this can be found in modern reverse engineering literature
(e.g., \cite{CarterFerranteThomborson:2003}, \cite{PandeRyder:1993}, 
\cite{MuthDebray:2000}). This work follows a new research
direction integrating hardness results to the software engineering 
and reverse engineering disciplines to the design of software 
protection systems (cf. \cite{WangEtal01},
\cite{ChowGuJohnsonZakharov:2001},
\cite{CarterFerranteThomborson:2003}).

\end{subsection}
\begin{subsection}{Prior art}\label{priorart}
In the past two decades (`80,`90) countering reverse engineering was 
endeavored either by encrypting the software, or by having it run  
on trusted environments or tamper-proof hardware devices.   
Unfortunately, typical hardware solutions have remained useless  or
too expensive 
(see, e.g., \cite{AndersonKuhn:1997:low}, \cite{magicmarker}, 
\cite{AppelGovindavajhala}); 
the case of typical software-only solutions has remained
invariably insecure (see, e.g., \cite{gosler85}, \cite{vanOorschot:2003}).
These solutions either violate Kherchoff's principle (e.g., rely on security by
obscurity) or at least rely on hypotheses on debuggers or hardware 
devices that cannot be verified (cf. \cite{MicaliReyzin03}).
``Computation with encrypted data''  (e.g., \cite{sander98towards}) provides 
an alternative approach, but no solution applicable to generic software has 
been found yet. 
So it has often been possible to circumvent anti-piracy 
procedures by reusable methods that could be wide-spread 
over the Internet\footnote{Crackers and hackers web-pages as 
\cite{Fravia} or \cite{reverseEngeneering} contain very complete lists 
of cracks and serial number generators for almost every licensed software.}.

Some novel software protection systems, such as our own, aim to 
discourage crackers by counter-attacking reverse engineering tools, 
for example \cite{WangEtal01} provides an obfuscation tool-set 
aimed at obstructing static (flow-sensitive) code analysis;
\cite{HorneEtal02} proposes tamper-proofing of license enforcement
by coupling license enforcement tools with  dynamically 
self-checking programs; \cite{ChowGuJohnsonZakharov:2001} 
presents an obfuscation method that aims at reducing de-obfuscation 
to solving the acceptance problem on linear Turing machines (which
is {\sf PSPACE}-complete). However, these models for security
are incomplete and there remains to answer if they are secure
in a realistic sense (e.g., cannot be countered by crackers).
The case of the Trusted Computing Group (\cite{tcg}) is different, 
TCG plans to provide a new generation of personal computers 
that, among other things, allow for DRM of content and software. 
This solution requires a technology that is now unavailable,
has not been inspected by the public, and it will take years before this 
solution is effectively implemented.

\end{subsection}
\end{section}
\begin{section}{Results}
This work addresses two complementary goals. First, to present a 
semi-automated method for software protection that addresses 
todays' problematic, and can be implemented and used within the actual 
technology. It enforces license policy, in the sense that it will not
run when policy is violated, and incorporates effectively traceable
fingerprints. 
As a second goal, we propose a very realistic model of security for software 
protection systems, and analyze the security of our system.

We will assume the reader has some knowledge of software development
and cryptography. An introduction to these subjects can be found in 
\cite{LippmanLajoie:1998} and \cite{vanstone96hac}.

\begin{subsection}{Architecture and implementation}
The protection system can be applied to the source code of a C/C++ program, 
developed for {\tt win32} or Unix platforms\footnote{The underlying method 
can be deduced from this paper and then be extended for protection systems
on Java, Python and other programming languages ---though that is not the 
aim of this paper.} 
requiring no additional hardware nor connectivity.
Additionally, we introduce a traitor-tracing system that detects these fingerprints
(e.g., if a licensee redistributes his copy on the internet, he can be traced by this
procedure). We describe both procedures and give implementation
directions.

The protection system requires the intervention of a programmer, which
need not be the original developer of the software, that we call 
{\em the developer}.
The protection process consists of two phases where the source code 
of a program is transformed into a protected build. On the first (manual) 
phase, the developer is required to add pragma directives (directives for short) 
to the source code of the program that is being protected ---specifying the location 
and name of the protection transforms (e.g., for fingerprinting or obfuscation). 
The manual phase is performed only once,  either during or after
development, and integrates to the development cycle enabling 
debugging and barely augmenting the development work. 
In the second (automated) phase, the protection system transforms the 
modified source code into a customized build according to both the 
directives added and a configuration file containing a license ID and 
license constraints. 

\end{subsection}
\begin{subsection}{Model and Security}
The threat model is defined by a developer that uses this protection
system to transform the source code of  {\em the program} into 
{\em (protected) builds}.
Each build includes an ID and license constraints, and is
delivered to the licensed customer as binary code. 
A valid attack against our system will consist on a sequence of
analyses and transformations to a set of builds done with crackers' 
tools. 

Crackers' tools are: Disassemblers and decompilers (e.g., \cite{IDA}), 
debuggers (e.g., \cite{ollydbg}), auto-decryptors and auto-decompressors, 
and analyzers for: API access patterns, application flow, 
and binary layout  data.
Theoretic counterparts for these tools are mainly static analysis 
methods (i.e., those analyzing the code from a frozen image of 
a build; see, e.g., \cite{Hetcht:1977}, \cite{CousotCousot:1977}), 
such as control- and data-flow static analysis; and dynamic analysis 
methods (i.e., those that infer the program's properties through 
running a copy; see, e.g., \cite{ball99concept}) such as
frequency spectrum and coverage concept analysis. 
We argue that by making our system invulnerable to these ``theoretic 
counterparts" we are in fact defending from the real attacks. 

Attackers are said to succeed at cracking our protection
system if they are able, either to bypass the license constraints,
or to erase the fingerprints (avoiding traitor tracing) using
the aforementioned tools.

\smallskip
The security provided by this system depends on both the program
being protected and the directives inserted during protection. This 
observation differences our method from most protection methods. 
Moreover, we cannot assert that programs protected by our system 
become un-crackable, but aim to prove that this system helps to 
make crackers job more difficult. 
Our method can profit from the syntax of source code implementations 
(e.g., flow chart, number of variables used, functions
implementations, etcetera) and the developer's ability. We shall 
give guidelines for the manual phase so that any programmer 
can use it.

As for the security brought by our system, first we remark that a new 
interpretation of the secure triggers (\cite{FuKaSaWa:2003}) technique, 
described in Section~\ref{triggers}, can be used to obfuscate programs 
securely (e.g., obfuscation is as secure as a block cipher is). 

Second, with our system the developer gains the ability to enforce
policy by binding the program's execution code with environment 
parameters.
In particular, by binding the program's execution with license 
information, the program will inject failures inside its own code, 
inevitably crashing, if there occur discrepancies between the 
expected values for these parameters and those actually assumed by them.
Failures are injected both stealthily and dynamically, making 
them more difficult to detect (e.g., by static analysis algorithms).

Third, the protection system embeds fingerprints during the protection 
process that are spread throughout the code producing customized 
builds. No two are alike (even locally). 
Further, the secure triggers technique referenced above will augment 
fingerprinting capabilities so that cracking triggers is necessary for 
cracking fingerprinting.

Finally, notice that \cite{barak01impossibility} does not apply to fingerprinting 
because our fingerprinting method does not pretend to hide fingerprints 
from crackers (as it happens with watermarking methods), but aims at 
replicating fingerprints throughout the complete build so they cannot be 
removed. 
In fact, this impossibility result does not apply to license enforcement
either, since in our case environment additional information ---that is 
independent of the obfuscation process--- needs to be fed for a 
protected program to run properly. 
Using a suitable strategy for the manual phase (see Section \ref{strategy}), 
both fingerprinting and license enforcement capabilities will change 
dynamically during the protected program's lifetime as the different branches
of the program are explored. 
More explicitly, the obfuscated portions of the program will be actually encrypted
with a block cipher (e.g., AES), and the keys needed for decryption
will be computed from environment variables and program parameters (some 
related to license conditions) on the fly as needed.
As a result, a cracker will not be able to assert whether he has completely 
cracked a build until he has decrypted every enciphered portion of code. 
We will argue later that this is a difficult job, and hence so is cracking our 
system.

\end{subsection}

\medskip
The paper continues as follows: in Section \ref{prelim} we isolate
the protection techniques required by this system. An implementation
 outline is drafted in Section \ref{truss}. Strategic considerations for 
 the manual phase follow in Section \ref{strategy}. 
Section \ref{analysis} includes our conclusions and discuss results.

\end{section}
\begin{section}{Tools and techniques for software protection}\label{prelim}
Our system performs transforms to the program following directives inserted 
in the manual phase. These transforms do not change the observable
behavior of the program. They comprise cryptographic or software 
engineering methods that we introduce here as stand-alone algorithms. 
Implementation details  follow in the next sections. 

\begin{subsection}{Obfuscation through secure triggers}\label{triggers}
Secure triggers are cryptographic primitives ideally
suited for solving malicious-host problems of mobile computing
(see \cite{FuKaSaWa:2003}, \cite{Waissbein:2004}, cf. \cite{Hohl:1998}).
Generally speaking, given a predicate (i.e., a binary valued function)
 $p:\{0,1\}^*\to\{\mathtt{true,false}\}$ 
and a secret procedure $f$, a cryptographic trigger is an algorithm
that executes the procedure $f$ on receiving the input $x$ only 
if $p(x)=\mathtt{true}$, else it returns nothing.  
Secure triggers encompass algorithms that compute functions 
$t(f,p)$ and are secure against white-box analysis, say, 
that given complete access to the algorithm that computes the function
$t(f,p)$, it is infeasible for an attacker to recover any semantic information 
regarding the procedure $f$. 

Every trigger's overall behavior is similar, after setup the trigger will
accept inputs and launch the secret procedure only if the trigger 
criterion is verified by the input. In \cite{FuKaSaWa:2003} three trigger 
examples: The {\sf simple trigger} will decrypt and launch
the secret functionality if the input received matches a predefined value;
the {\sf multi-strings trigger} decrypts and launches if it receives a sequence 
of values that contains a predetermined subsequence, and the
{\sf subset trigger} decrypts and launches only when certain specific 
bits of the input hold a predetermined value (see Appendix for details).
Browsing the code of these programs will render no key, nor what are
the {\em triggering} bits in the latter case.
The hardness result of Futoransky {\em et alii, ibidem}, states that: 
{\sl If there exists an ind-cpa secure block cipher (see \cite{vanstone96hac}), 
then no probabilistic polynomial time attacker can recover semantic 
information for $f$ when inspecting the algorithm $T(f,p)$ 
in any of the three examples described above.}
The appendix contains a description of these triggers and the
underlying security results. 

Let us now describe how is the simple trigger used in our protection
system. The use of other triggers can be derived from it.
Say that within the program's source code we isolated
a procedure $f$. Assume that the natural flow of the program 
assigns the value $\mathtt{k}$ to the local variable {\tt tmp}.
Let $E$ and $D$ denote a pair of symmetric encryption and
decryption primitives.
We then use a simple trigger ($p(x)=\mathtt{true}$ if, and only if, $x=\mathtt{k}$)
by replacing the procedure for $f$ by the algorithm on Figure \ref{fig1},
\begin{figure}[h]\label{fig1}
\begin{center}
\begin{minipage}{3in}
  {\em stored:} $\mathtt{iv},E(\mathtt{k,iv}),E(\mathtt{k},f)$.\\
  {\em input:} $\mathtt{tmp}$.\\
  {\em output:} $S$ or $\perp$.\\
  {\bf compute} $E(\mathtt{tmp,iv})$;\\
  {\bf If} $E(\mathtt{tmp,iv})= E(\mathtt{k,iv})$\\
 ${}{\qquad}${\bf then} \{{\bf output} $D(\mathtt{tmp},E(\mathtt{k},f))$\}; 
\end{minipage}
\end{center}
\caption{The simple trigger}
\end{figure}
where $E(\mathtt{k},f)$ denotes an encryption of a compiled $f$.
A best usage strategy and implementation details are included in 
Sections \ref{strategy} and \ref{analysis}.

\end{subsection}

\begin{subsection}{Fingerprinting}\label{finger}
We are particularly interested in software fingerprints that are 
robust and collusion resistant\footnote{A software fingerprint
is {\em robust} if the it remains even after disassembly, 
modification and reassembly.
The fingerprinting method is said {\em collusion-resistant},
if the method remains robust even when the attacker is
given a set of different fingerprinted builds but cannot produce
a single untraceable build.}. Our approach to fingerprinting
follows common practices (see, e.g.,  \cite{collberg-watermarking}, 
\cite{arboit-method}, \cite{wroblewski02general}) but profits from
our architecture design and the use of secure triggers.
Furthermore, the secure  triggers technique turns this difficult-to-defeat
fingerprinting system into a robust scheme.

The fingerprint module in our system is probabilistic: Every protected 
program is customized from a different ID (used as a seed). 
Static  fingerprints are embedded by random modifications on the 
syntactic structure or layout of the source code that do not modify its 
functionality. After compilation, this random changes remain in the build.
A suspected copy can be identified with the aid of our traitor tracing tool 
(see Section \ref{recog}) by different statistic correlation analyses 
between the suspected copy and those copies stored by the software
license owner.

The cornerstone of our fingerprinting method comes from the realization
that  programmers take arbitrary decisions during development, 
that this variations are present in the binaries, and that this slack 
can be harnessed for embedding fingerprints.
Our approach is to have the developer manually identify  the places in
the source code open to arbitrary decisions, and then have the fingerprinting
module to automatically randomize decisions on compilation.
For example,  developers arbitrarily decide the order in which a function 
accepts its arguments; with our method the developer will identify the 
arguments that can be arbitrarily reordered, then on each build the protection 
system will randomly reorder these arguments maintaining the code's logic.

Other ``permutables'' include: local and global variable definitions, function 
definitions, struct members, class data members, class methods, enumerated 
types, constant arrays, object code link order, if/then/else statements, independent 
statements. 
As a result, a single permutation will introduce multiple changes on several
parts of the binary code. For example, permuting the order of global 
variables definitions will produce a one byte difference in the binary code 
on each reference to these variables. 

Random generated implementations for common-use functions (e.g., 
manipulation of constants, multiplication of large integers, etcetera) 
provide yet another fingerprinting channel. Say, for example, numeric 
constants can be replaced, automatically in each build, by functions 
evaluating to those same constants ---with a minor performance penalty. 

Software byproducts, such as intermediate or temporary files, can also be 
fingerprinted by methods similar to those used above aiding forensic practices (see \cite{prosiseMandia}). Candidates for byproduct fingerprinting include 
intermediate or temporary files, saved documents,  configuration or 
state information, network traffic, and internal data structures. Fingerprints
can be embedded, e.g., as order permutations, formatting and document 
layout, and packet encapsulation.

Robustness is then achieved with no significant effort: 
since the marks are embedded in no particular portion of the build 
but distributed throughout the code. 
To erase these fingerprints an attacker would first need
to identify each of this random changes, then  disassemble the said
portion of the code and make the necessary modifications to erase
this random changes following the code's logic (e.g., if the protection
swaps a couple of global variables, a cracker attempting to erase
this mark would need to swap every appearance of this variables
on the build). 
A more thorough discussion follows in Section \ref{analysis}.
See also  e.g., \cite{BullTrevorsMaltonGodfrey02}
and \cite{marchionini95}. 

Dynamic fingerprints can also be inserted in several ways. For 
example, easter eggs (see, e.g., \cite{collberg-watermarking}) 
that {\em hatch} with fingerprinting information when accessed 
with special inputs can be embedded in the program and hidden
through the usage of secure triggers. 
Say, for example, that if someone inserts an entry to the program's 
database of a lady born on the year 2531, then the program decrypts 
and executes a function displaying the license ID on the screen.

\end{subsection}
\begin{subsection}{License Enforcement}\label{license}
We aim to make license enforcement by binding the program's 
(policy-conforming) execution to the values assumed for license 
parameters.
More explicitly, we establish two levels of protection. First 
level license checks are used to inform the licensee when 
the licensed program cannot be used at that moment (e.g., 
because it has expired).
No effort is made at this level neither to hide the location of these checks
in the program nor to prevent attackers from removing them\footnote{For 
example, if the software has an expiration date, then the protection scheme 
checks for the current time with a standard system call. A cracker can hook 
this system call always returning a date falling before expiration,
procedure completely breaks the first level of license enforcement.}.
The second level of  license enforcement does counter cracking.
Our parameter binding method consists in replacing certain program
constants by functions that evaluate to the expected value for these 
constants only when the license parameters hold values conforming 
with license policy.
The process is quite simple: During the manual phase of the protection, 
the developer identifies some constants in the source code that he wants 
to get bound on compilation, and he also supplies the functions  that 
return license parameters (e.g., the present time, or the host IP number). 
The system will automatically make the binding on the second phase.

We give a simplistic example for second level checks: 
Suppose that license policy establishes that the software expires on 
31-Dec-2009.
Then, the protected program will ``assume'' that the second (rightmost)
digit of the year is a $0$, and thus several appearances of the 
constant $0$ will be replaced by the variable ``second rightmost
digit of the year.'' If the year 2010 is reached, and the attacker
has circumvented the explicit checks (but not the parameter-binding
ones) every function that replaces $0$ by the variable ``second 
rightmost digit of the year'' will evaluate to $1$, injecting faults in
the program, and forcing the program to crash while failures spread 
during execution. 
In fact, different occurrences of the constant $0$ will be replaced by
different implementations of the function that returns the current time 
(e.g., the time of creation of temporary files and other timestamps within 
reach), making it more difficult  to thwart these checks.

More generally, execution and operational parameters can be used 
to specify license constraints such as 
number of records held on a database, the time of the creation of 
a record, the number of simultaneous users, usage time elapsed, 
and any machine identification parameter.

To circumvent this license enforcement method a cracker 
must identify every license check and swindle the 
application with bogus information (e.g., hooking these 
checks and always returning the correct values). 
Since most of these checks will be obfuscated by secure triggers, 
the cracker will need to break the obfuscation scheme to remove
license constraints.

\end{subsection}
\end{section}
\begin{section}{Implementation}\label{truss}
We give details for implementing a system that performs the 
automated phase of the protection process for C application 
projects (e.g., software applications) under Microsoft Visual 
Studio for the win32 platform.

We shall require the use of different cryptographic 
primitives that can to be chosen by the developer: A 
symmetric cipher, say AES in CBC mode, a hash function, 
say  SHA-1, and a random pool (see, e.g., \cite{vanstone96hac}).

\begin{subsection}{Architecture}
The system consists of three procedures: the crypto pre-processor 
procedure (CPPP), the compiler, and the post-processor.  Additionaly, 
it requires a library including the functions underlying triggers (e.g., 
decryption and integrity checks).
The CPPP is the first module executed, it receives as input a modified
source code (e.g., that includes directives) and outputs a randomized 
source code.
At startup it initializes the random pool as seeded by the configuration file
(the pool's required size is proportional to the number of directives in the
source code). 
Then, the CPPP parses the source code individualizing the portions of 
code marked for transformation.

Subsequently the transforms are applied according the random pool
and the marking directives. This system enables for the use of many
transforms. In the following paragraphs we give a flavor of the transforms
supported by this system introducing fingerprinting, license enforcement
and obfuscation transforms.
We start with an example of the permutation transform which reorders 
variable assignments:
\begin{center}
\begin{minipage}{3in} 
{\tt
fingerprint\_begin\_permute; \\
${}{\qquad}$a = 5;\\
${}{\qquad}$b = 4;\\
${}{\qquad}$tmp = "hello";\\
fingerprint\_end\_permute;
}
\end{minipage}
\end{center}
The parser will identify the three assignments between the begin
and end clauses.
Reordering is done by first enumerating the permutable lines
and then applying a successive swaps to the enumerated lines.
Notice that these modifications do not change the program's functionality.

Another fingerprinting directive is prepended to if/then/else statements
as follows:
\begin{center}
\begin{minipage}{4in} 
{\tt fingerprint\_if;\\
if ($a < 22$) printf("yes"); else printf("no"); }
\end{minipage}
\end{center}
According to the next bit in the random pool this will leave the 
statement as it is or replace it by the equivalent statement
\begin{center}
\begin{minipage}{4in} 
{\tt if (!$a< 22$) printf("no"); else printf("yes");}
\end{minipage}
\end{center}

The {\tt fingerprint\_constant(value,size[,func=value2])} directive can be used in the declaration 
of constants for fingerprinting and license enforcement.
The CPPP will replace the declaration of the constant  prepended by this 
directive by randomly generated arithmetic expression of size {\tt size} evaluating
to {\tt value}\footnote{Compiler's optimization options can be switched not to 
destroy these expressions, and still optimize producing a small slowdown.}.
For example, the declaration {\tt a:=2;} could be replaced by {\tt a:=4 - (3*2) + 5 - 2 + 1;} 
which evaluates to $2$. The optional argument is used for license enforcement,
the function {\tt func}  returns an operational parameter, that is assumed to take
the value {\tt value2} (see Section \ref{license} for details).

Triggers are handled by different encryption directives. The code to be encrypted
is enclosed between directives, say between  {\tt simple\_Trig\-ger\_be\-gin(key)}
and {\tt simple\_Trig\-ger\_end}.
The CPPP will identify each trigger occurrence and add calls to the corresponding 
trigger function (e.g., that of Figure \ref{fig1} in the case of the simple trigger) including
also integrity checks, and  will create a file in its working directory which contains 
references to the blocks associated to triggers (by specifying the starting line and 
ending line of each block in source code file).  
When running a protected program, if a function inside an encrypted block is 
required by the control flow of the program, the protected program will automatically 
compute the decryption key, (implicitly) check the validity of this key, decrypts 
the block, checks the block's integrity against a pre-stored hash value, and 
finally execute it.
The CPPP  also configures the project makefile (i.e., where ``details for files, 
dependencies and rules by which an executable application is built'' are
stored in msdev)  to link the post-processor and the additional library to the  project.

Then, the randomized source code computed by CPPP is 
compiled into a binary build with the msdev C compiler.
The output of this procedure is an executable binary, except no blocks are 
encrypted. Encryption is handled by the post-processor module.

Finally, the post-processor modifies the binary files computed by the compiler, 
by encrypting specified blocks as needed and completing parameters used by 
the triggers.
Explicitly, the post-processor makes two passes on these binary files. 
For each block marked for encryption, it first computes and stores its length 
in bytes and the symmetric key, and the hash value
for the cleartext; then, on the second pass, it encrypts the block and copies 
on the build overwriting cleartexts and inserting auxiliary information.

\end{subsection}
\begin{subsection}{Traitor tracing module}\label{recog}
The traitor tracing module detects static fingerprints using simple statistics
methods. Let $n$ be a positive integer chosen by the developer. Say, 
$n=10$. For every software build that has been delivered to a licensee, 
the traitor-tracing module will analyze the binaries making a dictionary of all the $n$ 
byte strings appearing within this file. 
This dictionary is constructed by hash tables  requiring a 
computation time proportional to the size of the program and the number
of copies delivered.  

When a suspected build is found it is parsed into $n$ bytes strings and each
string is looked up in the dictionary.  Then, for each protected build, the 
following statistic parameters are computed:  
i) the number of strings that can be found only in the suspected copy and this 
build, 
ii) it also computes the number of strings found in the suspected copy, this
build, and some other delivered build;
iii) the number of strings that can be found in the suspected copy and every
delivered build.

Dynamic fingerprints such as easter eggs can be detected automatically:  A 
single-sing-on procedure will start the program and take the necessary steps to 
insert the special entries on the databases so that the hidden license ID value
is displayed. 
More information on dynamic fingerprints can be found in Sections \ref{finger} 
and \ref{analysis}.

 \end{subsection}

\end{section}
\begin{section}{The developer's strategy}\label{strategy}
In this section we shall describe strategies for the manual phase
that will result in a stronger protection.
We remark that the security of programs protected by our system will 
depend on their characteristics and on the developer's job 
during the manual phase.

The recommended strategy will be aimed at making the fingerprinting
robust and making license checks hard to crack. We recommend to:
i) use the fingerprinting commands whenever possible, maximizing the
randomization within protected builds;
ii)  replace constants in the program with the license enforcement 
primitive in error-prone places of the program, so that when license 
fails the failures injected are hard to reproduce and will get the program 
to behave erratically;   
iii) spread license checks through the length of the code and
within every trigger's encrypted portion of software;  
iv) maximize the number of access channels to the license parameters 
(e.g., retrieve the actual time using different functions);
v) insert {\em fake} triggers, i.e., that are not reached by normal executions,
both allowing fingerprinting and making infeasible to the attacker the job 
of decrypting every trigger;
vi) nest occurrences of triggers;
vii) make the keys used by triggers non-obvious deriving them from variables 
that are permanently updated (e.g., so that these variables take the value of the key 
only when it is needed).

\end{section}
\begin{section}{Discussion}\label{analysis}
This section servers two purposes, on the one side it describes the strength of
this protection system, on the other it complements Sections \ref{prelim} and 
\ref{strategy} giving insight on how to counter the attackers' tools.
We analyze the strength of several attack strategies against our method.

We first discuss the effectivity of our traitor tracing module. Suppose that we 
have recovered a pirate copy and want to identify its origin.
Initially, the developer will try to check for explicit client IDs. In case they were 
removed by crackers, the traitor tracing module is run.

Attempts at ``destroying'' every fingerprint through code 
re-op\-ti\-mi\-za\-tion are futile for several reasons.
For example, because  optimization tools will not be able to
difference between low-use and fake code  ---being dead-code elimination
an intractable problem (\cite{Gaissarian:2000}, cf. \cite{bodik97partial}).

Frequency spectrum analysis and other kinds of dynamic analyses 
will not be able to difference easter-egg dynamic 
fingerprints from those pieces of code that are rarely used (e.g., code 
that is executed only under very particular situations). 
Also, since static fingerprints were provoked by 
arbitrary decisions (that might look meaningful inside the code) an 
automatic tool may not be able to remove them all. It turns out, that 
typical software solutions for re-optimization cannot be used to delete 
these fingerprints (we give more details below).

Furthermore, since certain parts of the code are encrypted by trigger 
procedures, the attacker will need to decrypt them to find out if they 
contain  fingerprints and have them deleted. 

As an experiment,  a $1.2\, Mb$ win32 executable was marked with 
fingerprinting directives (and no encryption directives). Twelve 
different builds were compiled (using different IDs) rising the 
compilation time of 45 minutes (without protection) in a couple
of minutes (and less than an hour in all if triggers are used).
The resulting files were analyzed using the traitor tracing tool 
doing statistics with $n=10$ bytes strings in a few seconds.  
The size of the resulting dictionary was of 440,000 values, for the 1,200,000 
totality of strings (recall that the file is 1.2Mb long). About 
182,000 of the values of the dictionary were present in every build.  For each
build, the percentage of strings appearing in it and no other
build ranged from 20\%\ to 30\% of the size of the dictionary ---giving an excellent 
identification ratio.  On the other hand, an average less than 40\% of the entries 
in each build were present on only one other build.  
Cut-and-paste attacks, collusion attacks attempting to replace pieces of a build 
by other build's  pieces in order to remove fingerprints, are not likely to succeed 
facing this statistics (e.g., almost half of the builds are needed to produce an
unmarked copy).
In fact, cut-and-paste attacks fail because the mixing builds produces 
inconsistencies in the variable's assignations, and in particular  for those 
used by triggers' functions.

\smallskip
To counter license constraints the attacker has two approaches (\cite{HachezEtal00})
he can either identify every function (within the protected software) executing a 
license check and patch it, or he can identify every attribute that is checked and patch
 it so the miss-match cannot be detected. In any case, attackers would need to analyze 
 the program's code in order to identify what is checked or how it is checked. 
Notice that static analysis tools will fail since  encrypted portions of code will not 
be readable by these tools. 

Also notice that, since the different portions of the code are disclosed gradually 
(e.g., the program is not decrypted at once), new checks might appear at any 
time (unannounced). An attacker cannot ensure he has removed every check
 unless every single portion of the code has been analyzed (even fake triggers). 
 
So far we have argued that cracking is impossible, unless every trigger occurrence
is inspected and license checks are thwarted. A method to systematically do this
would inevitably take the following steps:
i) Find the decryption algorithm entry points  (e.g., searching for decryption algorithm's
magic constants).
ii) Then modify the decryption function to include a key-logger by adding a procedure 
that saves the keys used whenever a portion of code is decrypted.
iii) Trigger every block, by intensively using the application exploring every 
possible execution path. This procedure will get every encrypted portion of
code decrypted.
Except step iii), all the others can be automated and efficiently implemented. 
However, following every path of a computer program is an intractable problem 
(e.g., as difficult as the halting problem)
that grows exponentially with the number of branches in the program.

As we noted, none of the above attacks, attempting to reveal every trigger 
block, is feasible. Further, a cracker will not be able to assert if a trigger is 
fake unless he understands all the related code.  
But, that is precisely our goal: making it necessary for a successful attack to take the
time and effort to understand the complete code of the program and then 
 erase the protection's fingerprints and thwart license constraints.

\end{section}



\begin{appendix}
\begin{section}{Secure triggers}
We follow \cite{FuKaSaWa:2003}. Futoransky {\em et alii, ibidem } describes 
different secure triggers in the ``universally composable security (UCS) 
framework'' of R.~Canetti  (\cite{Canetti:2001:UCS}). We give a concise 
description of the underlying
algorithms and security results (for a complete description see 
\cite{FuKaSaWa:2003} and \cite{Canetti:2001:UCS}).

Let $(\mathtt{Gen,Enc,Dec})$  be a {\sf ind-cpa} secure symmetric cipher.

\begin{subsection}{The simple trigger protocol}
Let $S\in\{0,1\}^*$ be a secret procedure described as a string of bits.  Fix $k\in\Z$
a security parameter.
On set up we run the key generation 
algorithm $\mathtt{Gen}$ to produce a key $\mathtt{k}$
of size $k$ and arbitrarily choose a string $\mathtt{iv}\in\{0,1\}^k$. Then 
we compute $\mathtt{Enc}_\mathtt{k}(\mathtt{iv}),\mathtt{Enc}_\mathtt{k}(S)$ 
and initialize the algorithm in Figure \ref{fig1} with these three values.

Security follows from the indistinguishability property of the symmetric cipher (see 
\cite[Th. 3.1]{FuKaSaWa:2003}). Or, in the language of \cite{GoldreichGoldwasserMicali:1986},
for any attacker $\mathcal{A}$ against this scheme there exists an attacker 
$\mathcal{A}^\prime$ against the {\sf ind-cpa} security of the block cipher
(with a single known plaintext), such that  the advantage $\mathcal{A}$ 
has is smaller than the advantage of $\mathcal{A}^\prime$.

\end{subsection}
\begin{subsection}{Subsequence trigger}
For the subsequence trigger procedure, the trigger criterion is satisfied
when a pre-defined subset of an input message (considered as a sequence
of bits) matches a particular value. Formally, let $s,k$ be positive integers 
with $s>k$. This trigger family is defined by the predicates
\begin{align*}
\Big\{p_K:\{0,1\}^s\to\{\mathtt{true,false}\};  K\subset\{1,2,\ldots,s\}\times\{0,1\},\#K=k,
\qquad\\
\mbox{ and if }(i,b),(i,b^\prime)\in K\mbox{ then }b=b^\prime\Big\}
\end{align*}
A predicate $p_K$ evaluates to true on  input 
$x=(x_1,\ldots,x_s)\in\{0,1\}^s$, if and  only if, for every pair 
$(i,b)$ in $K$, it holds true that $x_i=b$. 

To implement this trigger  we construct an auxiliary family
of (pol\-y\-no\-mi\-ally-computable, uninvertible) functions from 
$\{0,1\}^s$ to $\{0,1\}^k$, such that a member $\tau$ verifies: 
\begin{enumerate}
\item given $x$ in $\{0,1\}^s$, there exist 
indices $j_1,\ldots,j_k\in\{1,2,\ldots,s\}\subset\Z$ such that 
\subitem $\tau(x)=\tau(x_1,\ldots,x_s)=(x_{j_1},\ldots,x_{j_k})$, and
\subitem if $y\in\{0,1\}^s$ is such that $y_{j_\ell}=x_{j_\ell}$ for $1\le\ell\le k$,
then both values have the same output $\tau(x)=\tau(y)$.
\item $\tau$ is onto, and for every $y\in\{0,1\}^k$ the cardinality
of the preimage $\tau^{-1}(y)$ is $2^{s-k}$.
\end{enumerate}
Assume $\tau$ has this properties, and fix
values $x\in\{0,1\}^s$ and $\mathtt{b}:=(\mathtt{b}_1,\ldots,\mathtt{b}_k):=\tau(x)$.
Let $p$ be the predicate defined by $p(y)=\mathtt{true}$ if and only if
$\tau(x)=\mathtt{b}$.
By hypotheses, there exist indices $1\le i_1,\ldots,i_k\le s$ such that 
$\mathtt{b}=\tau(x)=(x_{i_1},\ldots,x_{i_k})$ and for every $y\in\{0,1\}^s$
that, for $1\le j\le k$, satisfies $y_{i_j}=x_{i_j}$, the equality
$\tau(x)=\tau(y)$ holds. 
Hence, if $p(x)$ is $\mathtt{true}$, then $p(y)$ is also true.

With out further ado let $Hash:\{0,1\}^*\to\{0,1\}^m$ denote a one-way hash function
and let the function family
$$
\left\{ \sigma_{(t_1,\dots,t_s)}:\{1,2,\ldots,s\}\times\{0,1\}^s\to\{0,1\}^k;
 t_i \in \{0,1\}^m,\mbox{ for } 1\le i\le s \right\}
 $$
be defined by the assignment $\sigma_{(t_1,\ldots,t_s)}(i,x):=y:=(y_1,\ldots,y_k)$ 
and the procedure in Figure \ref{fig2}.
\begin{figure}[h]\label{fig2}
\begin{center}
\begin{minipage}{4in}
    Stored: $(t_1,\ldots,t_s)$.\\
    Input: $i,(x_1,\ldots,x_s)$.\\
    Output: $(y_1,\ldots,y_k)$.\\ \\
    {\bf set} $i_1:=i; y_1:=x_i; I:=\{i_1\}$;\\
    {\bf for} $n:=2$ {\bf to} $k$ {\bf do:} \{ \\
    ${}{\qquad}${\bf compute} $i:=\big(Hash(y_1\|\ldots\|y_{n-1}) \oplus t_{i_n}\big)$ mod $(s)$;\\
    ${}{\qquad}${\bf compute} $i:=i+\#\{j:j\in I\wedge j\le i\}; I:=I\cup\{i\}$;\\
    ${}{\qquad}${\bf set} $i_n:=i; y_n:=x_i;$ \\
    ${}{\qquad}$\}\\
    {\bf output} $(y_1,\ldots,y_k)$;
\end{minipage}
\end{center}
\caption{The auxiliary function}
\end{figure}
Given any function $\sigma$ from this family, notice that for every $i,1\le i\le s$,
$\tau:=\sigma(i,\;):\{0,1\}^s\to\{0,1\}^k$ 
trivially verifies properties i) and ii). 

The algorithm for this trigger can now be explained. Let $S\in\{0,1\}^*$ be 
the secret procedure. Let $(\mathtt{Gen,Enc,Dec})$ be a {\sf ind-cpa} secure 
symmetric cipher. On the initialization, we run the key generation algorithm 
and get a key $\mathtt{b}=(b_1,\ldots,b_k)$ of size $k$,  we randomly chooses 
bit-strings $t_1,\ldots,t_s\in\{0,1\}^m$ of size $m$, and arbitrarily chooses a 
bit-string $\mathtt{iv}$ (of size $k$), finally we compute $\mathtt{Enc}_\mathtt{b}(\mathtt{iv}),\mathtt{Enc}_\mathtt{b}(S)$
and store these values.

Let $\sigma:=\sigma_{(t_1,\ldots,t_s)}$ be the function induced by the
stored values $t_1,\ldots,\linebreak t_s$. Let $x$ denote the input of the trigger 
algorithm, then this algorithm for every $i,1\le i\le s$ computes 
$\tau(i,x)$ and checks if 
$\mathtt{Dec}_{\sigma(i,x)}(\mathtt{Enc}_\mathtt{b}(\mathtt{iv}))$
for any $i,1\le i\le s$. If this happens, it must be that $\sigma(i,x)=\mathtt{b}$ 
and the algorithm (computes and) outputs the secret $S$.

Security follows from \cite[Th. 3.2]{FuKaSaWa:2003}. 

\end{subsection}
\begin{subsection}{Multiple-strings trigger}
Let $k,s\in\Z$ be integers with $s\ge 2$, where $k$ is the security
parameter and $s$ is the number of keys that are used to trigger.
The trigger family is then defined by the predicates
$$
\left\{p_{\mathtt{k}_1,\ldots,\mathtt{k}_s}:\{0,1\}^*\to\{\mathtt{true,false}\}; 
\mathtt{k}_1,\ldots,\mathtt{k}_s\in\{0,1\}^k\}\right\},
$$ 
where the predicate $p_{\mathtt{k}_1,\ldots,\mathtt{k}_s}(x)=\mathtt{true}$
on input $x$ if, writing $x=(x_1,\ldots,\linebreak x_t)$ there exist indices $i_1,\ldots, i_s$ 
such that $(x_{i_1},\ldots,x_{i_1+k-1})=\mathtt{k}_1,\ldots,(x_{i_s},\linebreak
\ldots,x_{i_s+k-1})=\mathtt{k}_s$.

We describe the algorithm for this trigger. Fix $s\in\N$. 
On initialization we use the key generation algorithm $\mathtt{Gen}$ to 
generate keys $\mathtt{k}_1,\ldots,\mathtt{k}_s$ of size $k$, compute a
random bit-string  $\mathtt{iv}$ of size $k$, and finally computes 
$\mathtt{Enc}_{\mathtt{k}_1}(\mathtt{iv}), \ldots,\mathtt{Enc}_{\mathtt{k}_s}(\mathtt{iv})$
and $\mathtt{Enc}_{\oplus_i\mathtt{k}_i}(S)$.  These values are stored
for the algorithm to access. 
For every input $x\in\{0,1\}^*$, the triggerer procedure checks for the existence 
of integers $1\le i_1,\ldots,i_k\le m$ such  that $\mathtt{Enc}_{(x_{i_j},\ldots,
x_{i_j+k-1})}(\mathtt{iv})=\mathtt{Enc}_{\mathtt{k}_j}(\mathtt{iv})$ holds
for all $j$. 
If it does, it then computes $\oplus_j(x_{i_j},\ldots,x_{i_j+k-1})$ and 
$S=\mathtt{Dec}_{\oplus_j{(x_{i_j},\ldots,x_{i_j+k-1})}}(\mathtt{Enc}_{\oplus_i\mathtt{k}_i}(S))$
and outputs $S$.

Security follows from \cite[Th. 3.3]{FuKaSaWa:2003}.

\end{subsection}
\end{section}
\end{appendix}

\end{document}